\documentstyle[prl,aps,twocolumn]{revtex}
\tighten

\begin{document}
\def\eqn#1{Eq.$\,$#1}
\def\mb#1{\setbox0=\hbox{$#1$}\kern-.025em\copy0\kern-\wd0
\kern-0.05em\copy0\kern-\wd0\kern-.025em\raise.0233em\box0}
\draft
\preprint{}
\title{Quasilinear theory of the 2D Euler equation}
\author{Pierre-Henri Chavanis}
\address{Laboratoire de Physique Quantique,
Universit\'e Paul Sabatier, 118 route de Narbonne 31062 Toulouse, France}
\date{\today}
\maketitle 

\begin{abstract}

We develop a quasilinear theory of the 2D Euler equation and derive an
integro-differential  equation for the evolution of the coarse-grained vorticity
$\overline{\omega}({\bf r},t)$. This equation respects all the invariance
properties of the Euler equation and conserves angular momentum in a circular
domain and linear impulse in a channel. We show under which hypothesis we can
derive a H-theorem for the Fermi-Dirac entropy and make the connection with
statistical theories of 2D turbulence.


\end{abstract}

Two-dimensional flows with high Reynolds numbers have the striking property of
organizing spontaneously into large-scale coherent vortices \cite{mcwilliams}.
The robustness of Jupiter's Great Red Spot, a huge vortex persisting for more
than three centuries in a turbulent shear between two zonal jets, is probably
related to this general phenomenon. Many other vortices are observed in
geophysical and astrophysical flows. Understanding the structure and formation
of these organized states is still a challenging problem.

To be explicit, we consider as an initial condition a stripe of uniform
vorticity $\omega=\sigma_{0}$ surrounded by irrotational flow $\omega=0$. This
stripe is unstable and generates a complicated mixing process leading to the
formation of a quasi-stationary vortex slightly diffusing with viscosity. This
is the classical shear layer (or Kelvin-Helmholtz) instability investigated
numerically in, e.g., \cite{ssr}. These authors propose to interprete the
quasi-equilibrium state as a state of maximum entropy under the constraint of a
fixed energy and circulation. This is motivated by the statistical theory of the
2D Euler equation developed by \cite{miller} Miller (1990) and \cite{rseq}
Robert \& Sommeria (1991). A coarse-graining procedure is introduced and a
mixing entropy is constructed to describe the chaotic interchange of vorticity
levels along the evolution. Since the vorticity levels cannot overlap, they
follow an exclusion principle and this leads to a statistics of a Fermi-Dirac
type. Comparision with numerical simulations \cite{ssr} shows a very good
agreement with the theoretical prediction in the core of the vortex  where the
fluctuations are sufficient to validate the ergodicity hypothesis. This
statistical mechanics of phase mixing is closely related to the theory of
``violent relaxation'' developed by \cite{lb} Lynden-Bell (1967) for
collisionless stellar systems (e.g., elliptical galaxies) described by the
Vlasov equation \cite{csr,cNY}.

Less is known concerning the relaxation towards equilibrium. This is clearly a
complicated task and analytical results will be obtained only by introducing
approximations. Our objective is to derive a kinetic equation respecting all the
conservation laws and invariance properties of the Euler equation and driving
the system towards the Fermi-Dirac state by increasing the mixing entropy. If
such a program can be realized this will  provide a useful subgrid scale model
allowing Large Eddy Simulations (L.E.S.) of 2D turbulence with potential
applications in geophysics \cite{prep}. A first step in this direction  was made
by \cite{rs} Robert \& Sommeria (1992) using a {\it variational} procedure. They
assumed that ``out of equilibrium, the system evolves so as to maximize the rate
of entropy production $\dot S$ while respecting all the constraints of the Euler
equation''. This Maximum Entropy Production Principle (MEPP) leads to an
equation for the coarse-grained vorticity of a generalized Fokker-Planck type
which can be compared succesfully with Direct Navier Stokes simulations
\cite{rs,rr}. Their method was extended by \cite{cs} Chavanis \& Sommeria (1997)
who derived a set of equations respecting, in addition, the invariance
properties of the Euler equation. However, the MEPP is relatively {\it ad hoc}
and  {\it assumes} that the system evolves towards a maximum entropy state. In
this letter, we consider a completely different approach based on a {\it
perturbative} expansion of the Euler equation. This is the counterpart of the
quasilinear theory introduced in plasma physics and in stellar dynamics for the
Vlasov equation \cite{kp,sl,chavM}. Within some approximations, we derive a new
kinetic equation of a generalized Landau type for the coarse-grained vorticity
and  prove a H-theorem for the Fermi-Dirac entropy (instead of postulating it).
The results of the MEPP are recovered as an approximation of our model.

For a two-dimensional incompressible and inviscid flow, the Euler equation can
be written:
\begin{equation}
{\partial\omega\over\partial t}+{\bf u}\nabla\omega=0 
\label{euler}
\end{equation}
\begin{equation}
{\bf u}=-{\bf z}\wedge\nabla\psi\qquad \qquad \omega=-\Delta\psi
\label{poisson}
\end{equation}
where $\omega {\bf z}=\nabla\wedge {\bf u}$ is the vorticity and $\psi$ the
streamfunction (${\bf z}$ is a unit vector normal to the flow). The velocity can
be expressed as an integral over the vorticity field as
\begin{equation}
{\bf u}({\bf r},t)=\int d^{2}{\bf r'}{\bf V}({\bf r'}\rightarrow {\bf
r})\omega({\bf r'},t)
\label{bs}
\end{equation}
where 
\begin{equation}
{\bf V}({\bf r'}\rightarrow {\bf r})=-{1\over 2\pi} {({\bf r'}-{\bf
r})_{\perp}\over |{\bf r'}-{\bf r}|^{2}}+{\bf V}_{b}({\bf r'}\rightarrow {\bf
r})
\label{kernel}
\end{equation}
represents the velocity created in ${\bf r}$ by a vortex of unit circulation
located in ${\bf r}'$ (${\bf r}_{\perp}$ is the vector ${\bf r}$ rotated by
$+{\pi\over 2}$). The term ${\bf V}_{b}({\bf r'}\rightarrow {\bf r})$ accounts
for boundary effects (${\bf V}_{b}={\bf 0}$ in an infinite domain) and can be
calculated with the method of ``images''.

In the situation described previously, the Euler equation builds up an intricate
filamentation at smaller and smaller scales. If we subdivise our domain into a
lattice of macrocells of size $\epsilon$, only the ``coarse-grained'' vorticity
$\overline{\omega}({\bf r},t)$ can reach a stationary state. This
``coarse-grained'' vorticity is defined by a double averaging process \cite{sl}:
a space average over the cell of size $\epsilon^{2}$ centered on ${\bf r}^{(i)}$
and a statistical average to express our ignorance of the precise manner in
which the phase filaments of vorticity $\sigma_{0}$ are distributed in the
macrocell: 
\begin{equation}
\overline{\omega}({\bf r}^{(i)})=\biggl\langle {1\over
\epsilon^{2}}\int_{\epsilon} \omega({\bf r}^{(i)}+{\bf r}')d^{2}{\bf
r}'\biggr\rangle
\label{cg}
\end{equation}
The fluctuating vorticity $\tilde\omega=\omega-\overline{\omega}$, satisfying
$\overline{\tilde\omega}=0$, is simply the difference between the exact
vorticity and the smoothed-out vorticity. The passage from discrete to
continuous variables for $\overline{\omega}$ requires a hypothesis of scale
separation. We shall assume that the velocity field ${\bf u}({\bf r},t)$
consists of a strong large-scale component and a weak small-scale component such
that the characteristic scale of $\overline{\bf u}$ is much greater than that of
$\tilde {\bf u}$ (we also assume that $\tilde\omega\sim\overline{\omega}$). This
hypothesis of scale separation was previously made by \cite{dub} Dubrulle \&
Nazarenko (1997). This is of course an idealization since the energy spectrum
never presents a clear-cut gap in practice. However, this approximation should
reasonably well  account for the nonlocal interactions between large eddies and
small scales fluctuations in 2D turbulence. A more general study relaxing this
hypothesis can be found in \cite{laval}.

If we take the local average of
the Euler equation (\ref{euler}), we obtain a convection-diffusion equation:
\begin{equation}
{\partial\overline{\omega}\over\partial t}+\overline{{\bf
u}}\nabla\overline{\omega}=-\nabla {\bf J}
\label{eulerlarge}
\end{equation}
for the coarse-grained field with a current ${\bf
J}=\overline{\tilde\omega\tilde{\bf u}}$ related to the correlations of the
fine-grained fluctuations. In turn, the fluctuations depend on the smoothed-out
field according to the equation:
\begin{equation}
{\partial\tilde{\omega}\over\partial t}+\overline{{\bf
u}}\nabla\tilde{\omega}=-\tilde {\bf u}\nabla\overline{\omega}-\tilde{\bf
u}\nabla\tilde\omega+\overline{\tilde{\bf u}\nabla\tilde\omega}
\label{eulersmall}
\end{equation}
obtained by substracting (\ref{euler}) and (\ref{eulerlarge}). Within the scale
separation hypothesis, we can neglect the non linear terms $\tilde{\bf
u}\nabla\tilde\omega$ and $\overline{\tilde{\bf u}\nabla\tilde\omega}$ which
represent the interactions of the small turbulent scales among themselves
\cite{dub}. However, unlike \cite{dub}, we keep the linear term $\tilde {\bf
u}\nabla\overline{\omega}$ which takes into account the interactions between
small and large scales. Its order of magnitude $\overline{\omega}^{2}\epsilon/L$
(where $L\gg\epsilon$ is the domain size) is relatively small but this term has
a cumulative effect (see equation \ref{lambda}) giving rise to a diffusion
process. We consider therefore the coupled system
\begin{equation}
{\partial\overline{\omega}\over\partial
t}+{L}\overline{\omega}=-\nabla\overline{\tilde\omega\tilde{\bf u}}
\label{large}
\end{equation}
\begin{equation}
{\partial\tilde{\omega}\over\partial t}+{L}\tilde{\omega}=-\tilde{\bf
u}\nabla\overline{\omega}
\label{small}
\end{equation}
where $L=\overline{\bf u}\nabla$ is an advection operator. This ``quasilinear
approximation'' is standard in plasma physics and in stellar dynamics for the
Vlasov-Poisson system \cite{kp,sl,chavM} but, to our knowledge,  it has never
been applied to the 2D Euler system. Owing to the various approximations
introduced, this theory can describe only the late quiescent stages of the
relaxation when the fluctuations have weaken.

Introducing the Greenian:
\begin{equation}
G(t_{2},t_{1})\equiv \exp \Biggl \lbrace -\int_{t_{1}}^{t_{2}}dt L(t)\Biggr
\rbrace
\label{greenian}
\end{equation}
we can immediately write down a formal solution of (\ref{small}), namely:
\begin{eqnarray}
\tilde\omega ({\bf r},t)=G(t,0)\tilde\omega ({\bf r},0)\nonumber\\
-\int_{0}^{t}ds G(t,t-s)\tilde {\bf u}({\bf r},t-s)\nabla\overline{\omega}({\bf
r},t-s)
\label{formal}
\end{eqnarray}
Although very compact, this formal expression is in fact extremely complicated.
Indeed, all the difficulty is encapsulated in the Greenian $G(t,t-s)$ which
supposes that we can solve the smoothed out Lagrangien flow:
\begin{equation}
{d{\bf r}\over dt}=\overline{\bf u}({\bf r},t)
\label{lagrangian}
\end{equation}
between $t$ and $t-s$.

The objective now is to substitute the formal result (\ref{formal}) back into
(\ref{large}) and make some closure approximation in order to obtain a
self-consistant equation for $\overline{\omega}({\bf r},t)$. If the vorticity
were purely advected by the stochastic velocity field ${\bf u}$ (like a passive
scalar), the interaction (\ref{bs}) would be switched off and we would end up
with a diffusion equation for $\overline{\omega}$ with a diffusion coefficient
$D\sim {1\over 4}\tau\overline{\tilde{{\bf u}^{2}}}$ where $\tau$ is the
decorrelation time \cite{csr,rr}. However, in the case of the Euler equation,
the velocity fluctuations are induced by the fluctuations of the vorticity
itself according to:
\begin{equation}
\tilde{\bf u}({\bf r},t)=\lambda \int d^{2}{\bf r'}{\bf V}({\bf r'}\rightarrow
{\bf r})\tilde\omega({\bf r'},t)
\label{lambda}
\end{equation}
Therefore, considering (\ref{formal}) and (\ref{lambda}), we see that the
vorticity fluctuations $\tilde\omega ({\bf r},t)$ are given by an iterative
process: $\tilde\omega (t)$ depends on $\tilde {\bf u}(t-s)$ which itself
depends on $\tilde\omega (t-s)$ etc... Since $|\tilde{\bf u}|$, of order
$\overline{\omega}\epsilon$, is much smaller than $|\overline{\bf u}|$, of order
$L\overline{\omega}$, we can solve this problem perturbatively. This is the
equivalent of the ``weak coupling approximation'' in plasma physics
\cite{kp,sl,chavM}. For convenience, we have introduced a counting parameter
$\lambda$ in (\ref{lambda}) which will be set equal to one ultimately.  To order
$\lambda^{2}$, we obtain after some rearrangements:
\begin{eqnarray}
{\partial\overline{\omega}\over\partial
t}+{L}\overline{\omega}={\partial\over\partial r^{\mu}}\int_{0}^{t}ds\int
d^{2}{\bf r'}d^{2}{\bf r''} V^{\mu}({\bf r'}\rightarrow {\bf r})\nonumber\\
\times G'(t,t-s) G(t,t-s)\nonumber\\ \times \biggl\lbrace V^{\nu}({\bf
r''}\rightarrow {\bf r})\overline{\tilde\omega ({\bf r}',t-s)\tilde\omega({\bf
r}'',t-s)} {\partial\overline{\omega}\over\partial r^{\nu}}({\bf
r},t-s)\nonumber\\ +V^{\nu}({\bf r}''\rightarrow {\bf r}')
\overline{\tilde\omega({\bf r},t-s)\tilde\omega ({\bf
r}'',t-s)}{\partial\overline{\omega}\over\partial r'^{\nu}}({\bf r'},t-s)
\biggr\rbrace
\label{eq1}
\end{eqnarray}
In this expression, the Greenian $G$ refers to the fluid particle ${\bf r}(t)$
and the Greenian $G'$ to the fluid particle ${\bf r}'(t)$. The contribution
proportional to $\lambda$ (not written) can be calculated with  the assuption
that $\tilde\omega$ is purely advected by the large scale velocity, i.e
$\tilde{\bf u}\nabla\overline{\omega}$ is neglected in (\ref{small}). This is
the case considered by \cite{dub}. However, in this approximation the
coarse-grained enstrophy $\int\overline{\omega}^{2}d^{2}{\bf r}$ is conserved
\cite{dub} and no trend towards a self-organized state (e.g. maximum entropy or
minimum enstrophy state) is apparent. The exchange of enstrophy between small
and large scales (and also the source of entropy) corresponds to higher order
corrections in the equation for $\tilde\omega$. In this letter, we shall
consider exclusively the term of order $\lambda^{2}$ which accounts for a
diffusion process but we do not claim that the term of order $\lambda$ must be
necessarily discarded.

To close the system, it remains to evaluate the correlation function
$\overline{\tilde\omega ({\bf r},t)\tilde\omega({\bf r'},t)}$. We  shall assume
that the scale of the kinematic correlations is small with respect to the
coarse-graining mesh size and take:
\begin{equation}
\overline{\tilde\omega({\bf r},t)\tilde\omega({\bf
r'},t)}=\epsilon^{2}\delta({\bf r}-{\bf r}')\overline{\tilde\omega^{2}}({\bf
r},t)
\label{corr1}
\end{equation}
This assumption is consistent with our scale separation hypothesis and was made
previously by \cite{csr,rr} for the Euler equation and by \cite{kp,sl,chavM} in
plasma physics. Now:
\begin{equation}
\overline{\tilde\omega^{2}}=\overline{(\omega-\overline{\omega})^{2}}=
\overline{\omega^{2}}-\overline{\omega}^{2}
\label{id}
\end{equation}
For the case that we consider, the exact vorticity field $\omega$ can take only
two values $\omega=0$ and $\omega=\sigma_{0}$. This implies that
$\overline{\omega^{2}}=\overline{\sigma_{0}\times
\omega}=\sigma_{0}\overline{\omega}$ and therefore:
\begin{equation}
\overline{\tilde\omega({\bf r},t)\tilde\omega({\bf
r'},t)}=\epsilon^{2}\delta({\bf r}-{\bf
r}')\overline{\omega}(\sigma_{0}-\overline{\omega})
\label{corr2}
\end{equation} 
Substituting this expression in equation (\ref{eq1}) and carrying out the
integration on ${\bf r}''$, we obtain:
\begin{eqnarray}
{\partial\overline{\omega}\over\partial t}+\overline{\bf
u}\nabla\overline{\omega}=\epsilon^{2}{\partial\over\partial
r^{\mu}}\int_{0}^{t}ds\int d^{2}{\bf r}' V^{\mu}({\bf r}'\rightarrow {\bf
r})_{t}\nonumber\\ \times \biggl\lbrace V^{\nu}({\bf r}'\rightarrow {\bf
r})\overline{\omega}'(\sigma_{0}-\overline{\omega}'){\partial
\overline{\omega}\over\partial r^{\nu}}\nonumber\\ +V^{\nu}({\bf r}\rightarrow
{\bf r}')\overline{\omega}(\sigma_{0}-\overline{\omega}){\partial
\overline{\omega}'\over\partial r'^{\nu}}\biggr\rbrace_{t-s}
\label{eq2}
\end{eqnarray} 
We have written $\overline{\omega}'_{t-s}\equiv \overline{\omega}({\bf
r}'(t-s),t-s)$, $\overline{\omega}_{t-s}\equiv \overline{\omega}({\bf
r}(t-s),t-s)$, $  V^{\mu}({\bf r}'\rightarrow {\bf r})_{t}\equiv V^{\mu}({\bf
r}'(t)\rightarrow {\bf r}(t))$ and $  V^{\nu}({\bf r}'\rightarrow {\bf
r})_{t-s}\equiv V^{\nu}({\bf r}'(t-s)\rightarrow {\bf r}(t-s))$ where ${\bf
r}(t-s)$ is the position at time $t-s$ of the fluid particle located in ${\bf
r}={\bf r}(t)$ at time $t$. It is determined by the characteristics
(\ref{lagrangian}) of the smoothed-out Lagrangian flow.

Equation (\ref{eq2}) is a non Markovian integro-differential equation: the value
of $\overline{\omega}$ in ${\bf r}$ at time $t$ depends on the value of the {\it
whole} vorticity field  at {\it earlier times}. If the decorrelation time $\tau$
is short, we can make a Markov approximation and  simplify the foregoing
expression in
\begin{eqnarray}
{\partial\overline{\omega}\over\partial t}+\overline{\bf
u}\nabla\overline{\omega}={\epsilon^{2}\tau\over 2} {\partial\over\partial
r^{\mu}} \int d^{2}{\bf r}' V^{\mu}({\bf r}'\rightarrow {\bf r})\nonumber\\
\times \biggl\lbrace V^{\nu}({\bf r}'\rightarrow {\bf r})
\overline{\omega}'(\sigma_{0}-\overline{\omega}'){\partial
\overline{\omega}\over\partial r^{\nu}}\nonumber\\ + V^{\nu}({\bf r}\rightarrow
{\bf r}')\overline{\omega}
(\sigma_{0}-\overline{\omega}){\partial\overline{\omega}'\over\partial
r'^{\nu}}\biggr\rbrace 
\label{eq3}
\end{eqnarray}   
In the case of an infinite domain, ${\bf V}({\bf r}\rightarrow {\bf r}')=-{\bf
V}({\bf r}'\rightarrow {\bf r})$ and we have the further simplification
\begin{eqnarray}
{\partial\overline{\omega}\over\partial t}+\overline{\bf
u}\nabla\overline{\omega}={\epsilon^{2}\tau\over 8\pi^{2}}
{\partial\over\partial r^{\mu}} \int d^{2}{\bf r}'K^{\mu\nu}({\bf r}'-{\bf
r})\nonumber\\ \times
\biggl\lbrace\overline{\omega}'(\sigma_{0}-\overline{\omega}')
{\partial\overline{\omega}\over\partial r^{\nu}}- \overline{\omega}
(\sigma_{0}-\overline{\omega}){\partial\overline{\omega}'\over\partial
r'^{\nu}}\biggr\rbrace 
\label{eq4}
\end{eqnarray}   
where 
\begin{equation}
K^{\mu\nu}({\bf r}'-{\bf
r})={\xi_{\perp}^{\mu}\xi_{\perp}^{\nu}\over\xi^{4}}={\xi^{2}
\delta^{\mu\nu}-\xi^{\mu}\xi^{\nu}\over\xi^{4}}
\label{K}
\end{equation} 
and ${\mb\xi}={\bf r}'-{\bf r}$. The symmetrical form of this equation is of
course reminiscent of the generalized Landau equation in plasma physics obtained
with a quasilinear theory \cite{kp,sl,chavM}.

Introducing a tensor
\begin{equation}
D^{\mu\nu}= {\epsilon^{2}\tau\over 2} \int d^{2}{\bf r}' V^{\mu}({\bf
r}'\rightarrow {\bf r})V^{\nu}({\bf r}'\rightarrow {\bf
r})\overline{\omega}'(\sigma_{0}-\overline{\omega}')
\label{Dmunu}
\end{equation}   
and a vector
\begin{equation}
\eta^{\mu}={\epsilon^{2}\tau\over 2}\int d^{2}{\bf r}'V^{\mu}({\bf
r}'\rightarrow {\bf r})V^{\nu}({\bf r}\rightarrow {\bf
r}'){\partial\overline{\omega}'\over\partial r'^{\nu}}
\label{etamu}
\end{equation}   
equation (\ref{eq3}) can be rewritten in the more illuminating form:
\begin{equation}
{\partial\overline{\omega}\over\partial t}+\overline{\bf
u}\nabla\overline{\omega}={\partial\over\partial r^{\mu}}\biggl \lbrack
D^{\mu\nu}{\partial\overline{\omega}\over\partial
r^{\nu}}+\overline{\omega}(\sigma_{0}-\overline{\omega})\eta^{\mu}\biggr\rbrack
\label{FP}
\end{equation}  
This equation has the structure of a generalized Fokker-Planck equation with a
diffusion term and a drift term. The diffusion term corresponds to the turbulent
viscosity introduced {\it ad hoc} in most parametrizations of turbulence.
However, this term alone breaks the conservation laws of the Euler equation. The
present theory shows that an additional {\it drift term} must exist in order to
recover these properties. The drift is non linear in $\overline{\omega}$ so that
(\ref{FP}) is not, strictly speaking, a Fokker-Planck equation. This
nonlinearity accounts for the constraint $\overline{\omega}({\bf r},t)\le
\sigma_{0}$ imposed at any time by the conservation of the fine-grained
vorticity (see equation (\ref{euler})).

Equation (\ref{eq3}) respects the invariance properties of the 2D Euler equation
and has the same structure as equation (23) of \cite{cs} Chavanis \& Sommeria
(1997) derived on the basis of thermodynamical arguments. In their work, the
constraints of the Euler equation were satisfied with the aid of Lagrange
multipliers. In this new approach, the conservation laws follow naturally from
the symmetrical structure of equation (\ref{eq3}), as for the usual Landau
equation (the linear impulse $P=\int\omega y d^{2}{\bf r}$ and the angular
mometum $L=\int\omega r^{2}d^{2}{\bf r}$ play the role of the  impulse ${\bf
P}=\int f {\bf v} d^{3}{\bf v}$ and kinetic energy $K=\int f{v^{2}\over
2}d^{3}{\bf v}$ in plasma physics). This is more satisfying from a physical
point of view. Moreover, in the thermodynamical approach, the increase of
entropy is {\it  postulated} whereas in the present situation a H-theorem for
the Fermi-Dirac entropy
\begin{equation}
S=-\int \biggl\lbrace {\overline{\omega}\over \sigma_{0}}\ln
{\overline{\omega}\over \sigma_{0}}+\biggl ( 1- {\overline{\omega}\over
\sigma_{0}}\biggr )\ln \biggl (1-{\overline{\omega}\over \sigma_{0}}\biggr
)\biggr \rbrace d^{2}{\bf r}
\label{FD}
\end{equation}  
results immediately from equation (\ref{eq3}). This is proved by taking the time
derivative of (\ref{FD}), substituting for (\ref{eq3}), interchanging the dummy
variables ${\bf r}$ and ${\bf r}'$ and summing the two resulting expressions. Of
course, the increase of entropy is due to the coarse-graining procedure which
creates some irreversibility (the indetermination on the position of the
vorticity levels in a cell). The entropy $S\lbrack\omega\rbrack$ for the exact
vorticity $\omega$ is conserved by the Euler equation as the integral of any
function of  $\omega$.

It is remarkable that a quasilinear theory is sufficient to generate a turbulent
viscosity (but also a drift) and a source of entropy. We do not  necessarily
have to advocate  the non linear terms in (\ref{eulersmall}) to get these
properties. Note also that the entropy associated with the (coarse-grained)
Euler equation is the Fermi-Dirac entropy (\ref{FD}) in agreement with the works
of \cite{miller,rseq} at equilibrium. Unfortunately, equation (\ref{eq3}) does
not conserve energy exactly.  Therefore, the system will ultimately relax
towards the solution $\overline{\omega}=\sigma_{0}/(1+\lambda  {\rm
exp}(\alpha\sigma_{0} r^{2}))$  which is the maximum entropy state at fixed
circulation and angular momentum. This means that our approximations break down
at very late times.

A further connection with the statistical theory of 2D turbulence can be found.
Equation (\ref{eq3}) is an integro-differential equation whereas the equations
derived from the MEPP \cite{rs,csr,cs} are differential equations. The usual way
to transform an integro-differential equation into a differential equation is to
make a guess for the function $\overline{\omega}'$ appearing in the integral. It
makes sense to replace $\overline{\omega}'$ by its optimal value
$\overline{\omega}'=\sigma_{0}/(1+\lambda {\rm exp}(\beta\sigma_{0}\psi'))$
maximizing entropy at fixed energy and circulation. Substituting in
(\ref{Dmunu}) (\ref{etamu}) and making a ``local approximation'', we obtain
\begin{equation}
{\mb\eta}=D\beta\nabla\psi
\label{drift}
\end{equation} 
\begin{equation}
D={\tau\epsilon^{2}\over 8\pi}\ln\biggl ({L\over\epsilon}\biggr
)\overline{\omega}(\sigma_{0}-\overline{\omega})
\label{diffusion}
\end{equation} 
In equation (\ref{drift}), we recover the form of the drift derived by
\cite{cPRE} Chavanis (1998b) in a point vortex model. The drift coefficient can
be interpreted as an Einstein formula. Substituting for the drift in (\ref{FP})
we recover the equation 
\begin{equation}
{\partial \overline{\omega}\over \partial t}+\overline{\bf
u}\nabla\overline{\omega}=\nabla(D(\nabla\overline{\omega}+\beta
\overline{\omega}(\sigma_{0}-\overline{\omega})\nabla\psi))
\label{rs}
\end{equation} 
obtained by \cite{rs} Robert \& Sommeria (1992) using a Maximum Entropy
Production Principle. Equation (\ref{rs}) can be interpreted as a generalized
Fokker-Planck equation \cite {cPRE}. Note that the present approach provides the
value (\ref{diffusion}) of the diffusion coefficient which was left unknown by
the variational principle \cite{rs}. This value coincides with the estimate of
\cite{csr,rr}  based on a passive scalar model.  

The results of this letter can be extended to an arbitrary spectrum of vorticity
levels \cite{prep} for a wider class of initial conditions. These results also
complete the analogy between 2D turbulence and stellar systems investigated by
the author \cite{csr,cNY}.

{\it Acknowledgements:} I thank B. Dubrulle, J. Sommeria, R. Robert and U.
Frisch for their interest in this study.

\end{document}